# 18 GHz Solidly Mounted Resonator in Scandium Aluminum Nitride on SiO$_2$/Ta$_2$O$_5$ Bragg Reflector

Omar Barrera, Nishanth Ravi, Kapil Saha, Supratik Dasgupta, Joshua Campbell, Jack Kramer, Eugene Kwon, Tzu-Hsuan Hsu, Sinwoo Cho, Ian Anderson, Pietro Simeoni, Jue Hou, Matteo Rinaldi, Mark S. Goorsky, and Ruochen Lu*Abstract*—This work reports an acoustic solidly mounted resonator (SMR) at 18.64 GHz, among the highest operating frequencies reported. The device is built in scandium aluminum nitride (ScAlN) on top of silicon dioxide (SiO$_2$) and tantalum pentoxide (Ta$_2$O$_5$) Bragg reflectors on silicon (Si) wafer. The stack is analyzed with X-ray reflectivity (XRR) and high-resolution X-ray diffraction (HRXRD). The resonator shows a coupling coefficient ($k^2$) of 2.0%, high series quality factor ($Q_s$) of 156, shunt quality factor ($Q_p$) of 142, and maximum Bode quality factor ($Q_{max}$) of 210. The third-order harmonics at 59.64 GHz is also observed with $k^2$ around 0.6% and $Q$ around 40. Upon further development, the reported acoustic resonator platform can enable various front-end signal-processing functions, e.g., filters and oscillators, at future frequency range 3 (FR3) bands.

*Index Terms*—acoustic resonators, piezoelectric devices, solidly mounted resonators, thin-film devices## I. Introduction

PIEZOELECTRIC resonators have been used for radio frequency (RF) front-end applications, e.g., filters and oscillators. Piezoelectric devices transduce electrical signals into mechanical vibrations and process within the acoustic domain [1]. The main differentiator of acoustics over electromagnetic (EM) technology is that it provides four orders of magnitude smaller sizes along with better frequency selectivity [2]–[4]. Thus, thin-film bulk acoustic wave resonators (FBAR) and surface acoustic wave (SAW) devices are the dominant front-end filtering solutions [5]–[8].

More recently, the pursuit for faster data rate is driving wireless communication systems into higher frequency bands, e.g., frequency range 3 (FR3, 7.125 GHz to 24.25 GHz) and millimeter-wave bands (above 30 GHz) [9], [10]. The frequency scaling causes new challenges in acoustics as device performance degrades at higher frequencies [11], [12]. The main issue is that the acoustic wavelengths fall below 500 nm. One method is to use ultra-thin films or fine feature size electrodes [13]–[19], but the mechanical and electrical loss will be high, marked by reduced quality factor ($Q$). Alternatively, overmoding with higher-order tones in a larger cavity is

Manuscript received 1 June 2024; revised XX June 2024; accepted XX June 2024. This work was supported by DARPA COmpact Front-end Filters at the ElEment-level (COFFEE).
O. Barrera, J. Campbell, J. Kramer, T.-H. Hsu, S. Cho, I. Anderson, and R. Lu are with The University of Texas at Austin, Austin, TX, USA (email: omarb@utexas.edu). S. Dasgupta, and J. Hou are with Bühler Leybold Optics. K. Saha, P. Simeoni, and M. Rinaldi are with Northeastern University, Boston, MA, USA. N. Ravi, E. Kwon, and M. S. Goorsky are with University of California Los Angeles, Los Angeles, CA, USA.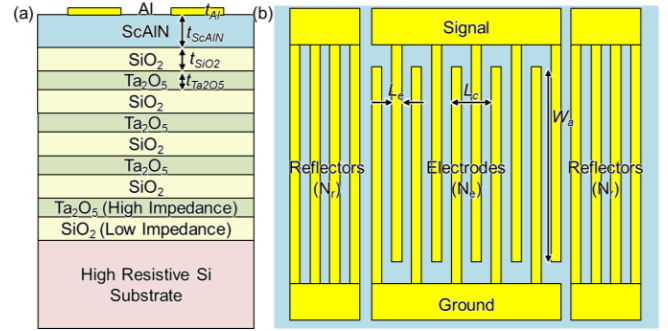

Fig. 1 Device schematics in (a) cross-sectional view of a unit cell, and (b) top view of the electrodes and reflectors.

Table I. Device Dimensions

| Sym. | Parameter | Value | Sym. | Parameter | Value |
|---|---|---|---|---|---|
| $t_{Al}$ | Al Thickness | 40 nm | $\Lambda$ | Cell length | 432 nm |
| $t_{ScAlN}$ | ScAlN Thickness | 68 nm | $W_e$ | Electrode Width | 108 nm |
| $t_{SiO2}$ | SiO$_2$ Thickness | 75 nm | $N$ | Cell Number | 50 |
| $t_{Ta2O5}$ | Ta$_2$O$_5$ Thickness | 65 nm | $N_r$ | Lateral Reflector Number | 15 |
| $N_{Bragg}$ | Bragg Reflector Pair Number | 8.5 | | | |

feasible, this approach has been demonstrated in single and periodically poled piezoelectric (P3F) devices [20]–[25]. However, this comes with fabrication complexities in P3F stacks, whereas single film suffers from low electromechanical coupling ($k^2$) due to charge cancellation.

Lately, the development of thin-film piezoelectric material, design, and fabrication has enabled a series of acoustic resonators beyond 18 GHz, mostly using suspended thin-film lithium niobate (LiNbO$_3$) [26]–[29] and scandium aluminum nitride/aluminum nitride (ScAlN/AlN) [21], [30]–[32]. Despite the remarkable advance of the state of the art, one issue remains for the power handling, which is intrinsically weak due to their structure with suspended membranes.

One promising platform is solidly mounted resonators (SMRs) [33]–[35]. In SMRs, the piezoelectric resonant cavity is above an acoustic quarter-wavelength Bragg reflector, consisting of alternating low and high acoustic impedance layers, deposited on the top of a carrier wafer [Fig. 1(a)]. The advantages of SMRs include intrinsically stable mechanical structure and high power handling [36]. However, scaling high $Q$ and $k^2$ SMRs beyond 10 GHz is not trivial, as the reflector also needs to be scaled, causing additional challenges, both on the design and microfabrication end. So far, SMRs are mostly

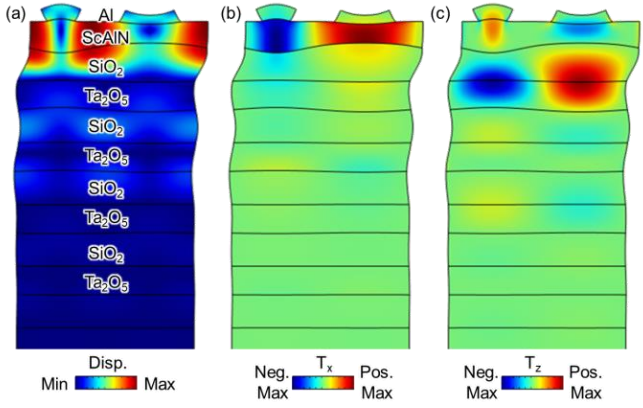

Fig. 2 Simulated acoustic mode shapes in (a) displacement, (b) $T_x$, and (c) $T_z$.

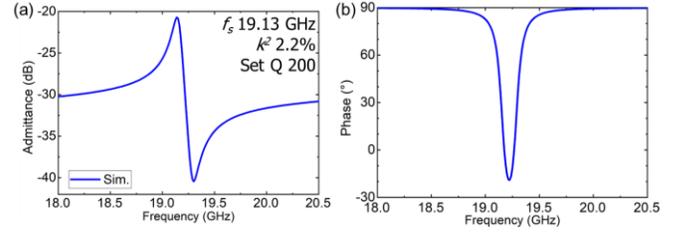

Fig. 3 Simulated frequency domain admittance in (a) magnitude and (b) phase.

Table II. Bragg Reflector Material Parameters in FEA

| Sym. | Parameter | SiO$_2$ | Ta$_2$O$_5$ |
|---|---|---|---|
| $\rho$ | Density | 2200 kg/m$^3$ | 6850 kg/m$^3$ |
| $E$ | Young's Modulus | 70 GPa | 162 GPa |
| $v$ | Poisson Ratio | 0.17 | 0.43 |
| $Z$ | Acoustic Impedance | 12.4 Mkg/(m$^2$·s) | 33.3 Mkg/(m$^2$·s) |
| $V$ | Acoustic Velocity | 5640 m/s | 4860 m/s |

limited to sub 10 GHz [37]–[43].

In this work, we report an SMR at 18.64 GHz using ScAlN on top of silicon dioxide (SiO$_2$) and tantalum pentoxide (Ta$_2$O$_5$) Bragg reflectors on silicon (Si) wafer. The resonator shows $k^2$ of 2.0%, high series quality factor ($Q_s$) of 156, shunt quality factor ($Q_p$) of 142, and maximum Bode quality factor ($Q_{max}$) of 210. The third-order harmonics at 59.64 GHz is also observed with $k^2$ around 0.6% and $Q$ around 40. The device also features a temperature coefficient of frequency (TCF) of −47.6 ppm/K for series resonance and −60.6 ppm/K for shunt resonance.

Upon further development, the reported acoustic resonator platform can enable various front-end signal-processing functions, e.g., filters and oscillators, at FR3 and mmWave bands.

## II. DESIGN AND SIMULATION

The stack of the proposed SMR is shown in Fig. 1 (a), with the key dimensions listed in Table I. The stack includes 40 nm aluminum (Al) electrode on the top of 68 nm Sc$_{0.3}$Al$_{0.7}$N, deposited on SiO$_2$/Ta$_2$O$_5$ Bragg reflector pairs on Si carrier wafer. The Bragg reflector consists of 8.5 pairs (17 layers in total) of thin films, with alternating 75 nm SiO$_2$ and 65 nm Ta$_2$O$_5$. The transducer design follows a conventional interdigitated electrodes (IDT) design with shorted metallic electrodes on the side [Fig. 1 (b)] [44]. The cell length (Λ) is chosen as 434 nm to excite a resonance around 18 GHz for the given Sc$_{0.3}$Al$_{0.7}$N thickness. The electrode width is 108 nm and is based on a fabrication process previously validated in [45]. The period of the lateral IDT is 432 nm, and the electrode width is 108 nm. The resonator layout includes 50 pairs of IDTs and 15 pairs of reflectors on each side.

In operation, the alternating electric field between IDT excites a confined longitudinal mode in ScAlN [displacement mode shape in Fig. 2 (a)]. The piezoelectric transduction is achieved from the thickness-direction electric field component, coupled into the lateral stress component [$T_x$ in Fig. 2 (b)] and thickness stress component [$T_z$ in Fig. 2 (c)], via piezoelectric coefficients $e_{31}$ and $e_{33}$, respectively. The piezoelectric coefficients follow that in [46]. In the thickness direction, the acoustic energy is confined, as the Bragg reflector transforms the impedance of the substrate to a minimal value comparable to air, effectively providing a free boundary condition at the bottom of ScAlN, supporting bulk acoustic modes on the top of the reflectors.

The stack is optimized via three-dimensional (3D) eigenmode analysis of a unit cell with one lateral wavelength in COMSOL finite element analysis (FEA). Periodic boundary conditions are applied on the sides. A perfectly matched layer (PML) is included at the bottom to represent the carrier Si wafer. The parameters of the Bragg reflector are listed in Table II, showing SiO$_2$ and Ta$_2$O$_5$ as low and high acoustic impedance layers, respectively. The key material properties are listed in Table II [38]. The acoustic impedance for longitudinal waves are 12.4 Mkg/(m$^2$·s) and 33.3 Mkg/(m$^2$·s) for SiO$_2$ and Ta$_2$O$_5$, respectively. The expected fractional bandwidth (FBW) for the Bragg reflection is [47] calculated to be:

$$FBW = \frac{4}{\pi} \cdot arcsin\left(\frac{Z_2 - Z_1}{Z_2 + Z_1}\right) \quad (1)$$

which yields FBW of 60.7%, indicating wideband function.

The Bragg reflector thickness is initially selected based on the longitudinal acoustic quarter wavelengths at 18 GHz, then optimized by parametrically sweeping the stack thickness, toward ensuring the eigenfrequency is purely real, indicating no energy loss into the substrate, modeled by the PML. The ScAlN thickness, Al thickness, and lateral wavelengths are selected with consideration of high $k^2$ at 18 GHz and also microfabrication limits. The mode shapes in Fig. 2 indicate that the vibration is well confined, and its amplitude exponentially decays into the Bragg reflector, validating the design.

The frequency domain simulated admittance of SMR is plotted in Fig. 3 (a) and (b) in magnitude and phase, respectively. The extracted $k^2$ is 2.2%, obtained via Butterworth-Van Dyke (BVD) fitting, which is equivalent to $k^2 = \pi^2/8 \cdot (f_p^2/f_s^2 - 1)$ for this case, without EM effects, e.g., routing inductance and resistance [20]. Mechanical $Q$ is set to 200 based on that reported in recent ScAlN works around 18 GHz [48]. The results show great promise for SMR at 18 GHz and beyond. Future enhancement of $k^2$ will require the implementation of a bottom electrode layer between the ScAlN and Bragg reflector, introducing additional fabrication complexity. Here, the feasibility of higher frequency SMR is showcased.

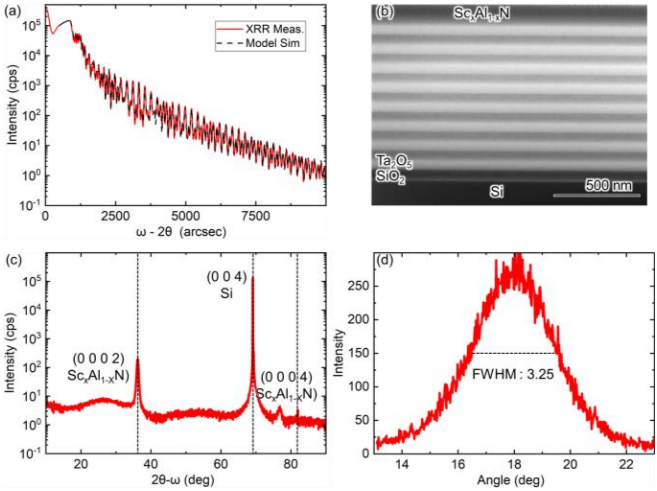

Fig. 4 (a) Measured XRR scan against layered model data, (b) Cross-section SEM of the stack and (c) Diffraction scan with detected material peaks (d) XRD rocking curve and FWHM estimation.

Table III. Comparison between Measured and Expected Stack Parameter

| Material | AlScN | SiO$_2$ | Ta$_2$O$_5$ |
|---|---|---|---|
| Average Density (%) | 87 | 71.8 | 91.8 |
| Expected Thickness (nm) | 67.6 | 75.2 | 65.0 |
| XRR Thickness (nm) | 81.7 | 74.3 | 65.0 |

## III. Fabrication

The fabrication starts with depositing the 8.5 pairs of SiO$_2$/Ta$_2$O$_5$ Bragg reflectors on the top of the Si wafer. The large number of pairs is chosen to validate the consistency and repeatability of the process. The deposition is done with a HELIOS 800 sputter coater by Buhler Leybold Optics, equipped with three magnetron sputter stations for physical vapor deposition (PVD), a plasma beam source (PBS) as an ion-assist source, two electrical heaters, and an optical monitoring system (OMS) for in-situ thickness monitoring. The alternating SiO$_2$ and Ta$_2$O$_5$ layers in the Bragg reflector stack were sputtered through the PVD stations using a plasma-assisted reactive magnetron sputtering (PARMS) [49] approach on Si-wafers. In PARMS, the PVD stations are equipped with metallic targets and operate with a mixed gas (Ar and O$_2$), while the process is controlled using lambda probes. The lambda probes help stabilize the process at an ideal working point where it operates at the edge of the oxidic limit without oxidizing the target surface, hence ensuring perfectly stoichiometric oxide layers with optimized deposition rates. The process is supported by the PBS, which provides oxygen radicals for better O$_2$ intercalation and reduces absorption in the films. The thickness was monitored in-situ by OMS in transmission with self-error compensation of optical thicknesses to achieve accurate physical thickness of the stack [50]. The coatings were performed at a deposition temperature of 150 °C.

Next, a 68 nm thin film of Sc$_{0.3}$Al$_{0.7}$N was deposited onto the top SiO$_2$ layer using an Evatec Clusterline-200 magnetron sputtering system. A 12-inch Sc$_{0.3}$Al$_{0.7}$N casted target has been utilized for this process. During the deposition, the target and substrate were positioned 20 mm apart. A steady flow of nitrogen gas (N$_2$) at a rate of 20 sccm was maintained, while the

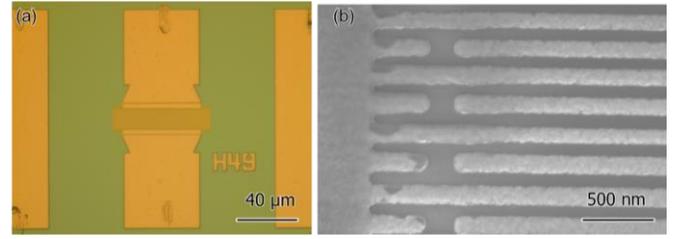

Fig. 5 (a) Optical and (b) SEM images of the fabricated resonator.

substrate was kept at a temperature of 400 °C. No argon was introduced to minimize roughness and abnormally oriented grains (AOG) formation on the film surface. 5 kW of pulsed-DC was applied to the target at 100 kHz with 88% on duty cycle for 103 seconds (0.66 nm/s deposition rate). Afterward, the top electrodes are patterned with electron-beam lithography, 40 nm Al IDT evaporation, and lift-off process. Finally, the busline regions are thickened by 300 nm evaporated Al.

The X-ray reflectivity (XRR) and high-resolution X-ray diffraction (HRXRD) measurements were performed on a D1 (Bruker/Jordan Valley) diffractometer using Cu Kα1 radiation with an incident beam optic to produce a parallel beam and a Si (220) monochromator. Specular XRR scans were performed using a 0.01 deg scattered beam slit and the HRXRD measurements used a 0.3 deg scattered beam slit. The XRR scans were simulated using REFS, a modeling program. The match between the experiment and the simulation confirms that the layer thickness is highly repeatable and uniform. The Scandium mole fraction in the Al$_{1-x}$Sc$_x$N alloy was determined through a combination of a symmetric 2θ-ω scan [(0002) and (0004) reflections] and an asymmetric 2θ-ω scan across the (10-13) reflection to measure the a and c lattice parameters. These were then matched to the lattice parameters in reference [51]. A limited area pole figure around the (10-13) reflection was also produced. In this case, a detector aperture of 1.7 deg was utilized. A scanning electron microscopy (SEM) image was produced after focused ion beam cutting to reveal the layer thicknesses.

Fig. 4 (a) shows the specular XRR scan and a simulated scan that superimposes the experimental data. For the simulated scan, the modeled structure included eight identical layers of Ta$_2$O$_5$ (thickness = 64.3 nm) and nine layers of SiO$_2$ (thickness 74.8 nm) as well as an AlScN layer approximately 81.6 nm thick. The fringe periodicity determines the period thickness, the equivalent narrowness of the peaks at low and higher angles indicates there is little thickness drift or dispersion and the relative amplitude of the peaks is primarily determined by the thickness ratio of the two layers. The Al$_{1-x}$Sc$_x$N composition predicted from the XRR simulation is approximately X(Sc) = 0.32 but the XRR measurement is not very sensitive to this composition. The comparison of measured and expected stack parameters are listed in Table III.

Fig. 4 (b) is the SEM image of the cross-section of layers. The Ta$_2$O$_5$ layers are lightest here, the SiO$_2$ layers darker, and the AlScN layer is the darkest of all. The uniform thickness of the layers and the number of layers modeled in the XRR result is consistent with the SEM image. Fig. 4 (c) shows a diffraction scan over a 2θ range from 10-90 deg. In addition to the (004) Si substrate peak, the peaks at 36.2 deg and 76.8 deg correspond to the (0002) and (0004) AlScN peaks, respectively. No other AlN peaks are observed, indicating that the AlN is highly oriented. A pole figure near the (10-13) expected reflection shows weak six-fold ordering. The composition of the Al$_{1-x}$Sc$_x$N was determined to be X(Sc) ~ 0.31-0.32. Fig 4(d) shows the measured rocking curves of the thin film, with a full width at half maximum (FWHM) of 3.25, comparable to those reported in recent works [30], [31].

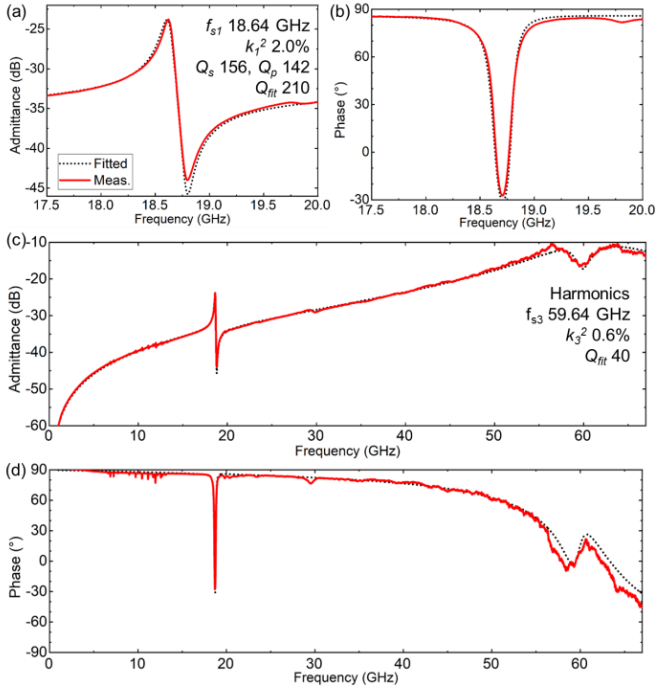

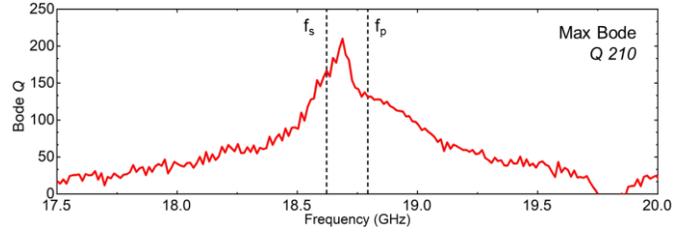

Fig. 8 Extracted Bode $Q$ at 18 GHz.

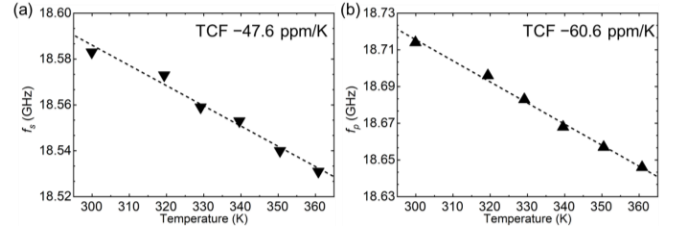

Fig. 9 Extracted TCF of the (a) series and (b) shunt resonances at 18 GHz.

Table V. Comparison to Prior SMR Work

| Ref. | Mirror Stack | $f_s$ (GHz) | Q | $k^2$ (%) | FoM |
|---|---|---|---|---|---|
| [36] | SiO$_2$/Ta | 9.5 | 400 | 2 | 8.0 |
| [37] | SiO$_2$/Ta$_2$O$_5$ | 3.5 | 225 | 20 | 45 |
| [38] | SiO$_2$/W | 7.5 | 2500 | N/A | N/A |
| [40] | N/A | 4.9 | 565 | 24 | 135 |
| [41] | SiO$_2$/AlN | 4.7 | 246 | 19.7 | 48.4 |
| [42] | Al/W | 55.7 | 95 | 3.5 | 3.3 |
| This Work | SiO$_2$/Ta$_2$O$_5$ | 18.6 | 210 | 2.0 | 4.2 |

Fig. 6 Measured acoustic resonator admittance and fitting results for zoom-in (a) magnitude, (b) phase, and wideband (c) amplitude, and (d) phase.

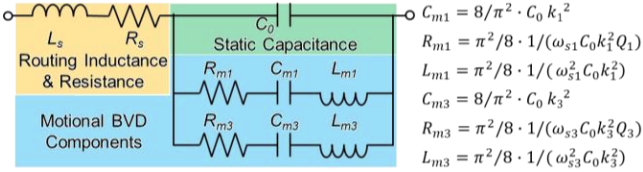

Fig. 7 Modified mmWave multi-branch BVD model for parameter extraction.

Table IV. Extracted Device Parameters

| Sym. | Parameter | Value | Sym. | Parameter | Value |
|---|---|---|---|---|---|
| $f_{s1}$ | Resonance | 18.64 GHz | $f_{s3}$ | Harmonics Resonance | 59.64 GHz |
| $k_1^2$ | Coupling | 2.0% | $k_3^2$ | Harmonics Coupling | 0.6% |
| Q | Fit Q | 210 | $Q_3$ | Harmonics Fit Q | 40 |
| $R_s$ | Series Resistance | 3.5 Ω | $L_s$ | Series Inductance | 0.04 nH |
| $C_0$ | Static Capacitance | 157 fF | | | |

Fig. 5 (a) and (b) show optical and scanning electron microscope (SEM) images of the fabricated resonators, respectively. The key dimensions are listed in Table I.

## IV. MEASUREMENT

The resonators are first measured using a Keysight vector network analyzer (VNA) in air at −15 dBm power level. Both zoom-in and wideband admittance and phase of the resonators are plotted in Fig. 6 (a)-(d), fitted with the mmWave MBVD circuit model [26] in Fig 7. The parameters for fitting are listed in Table IV. Unlike conventional MBVD models, the inductive effects from routing inductance $L_s$ are included together with $R_s$. Such $L_s$ and static capacitance $C_0$ forms an EM resonance at higher frequencies, indicated by the inductive phase beyond 55 GHz. Two motional branches are included for the 18 GHz and 59 GHz tones, respectively, with corresponding motional elements $L_m$, $C_m$, and $R_m$.

The resonance at 18.64 GHz show shows $k^2$ of 2.0%, high series quality factor $Q_s$ of 156 and shunt quality factor $Q_p$ of 142. The harmonics at 59.64 GHz is also observed with $k^2$ around 0.6% and extracted $Q$ around 40. The Bode $Q$ of the first tone is plotted in Fig. 8, showing maximum Bode quality factor $Q_{max}$ of 210 [52].

Next, TCF is measured with a Lakeshore cryogenic probe station TTPX, where the temperature varies between 300 K and 360 K. The resonances drift to lower frequencies at elevated temperature, due to negative TCF. The series and shunt TCF, defined by the frequencies where the admittance is pure real, are measured and fitted in Fig. 9 (a) and (b), respectively. The TCF is −47.6 ppm/K for series resonance and −60.6 ppm/K for shunt resonance. The value is higher than that in reported suspended ScAlN/AlN resonators [53], likely impacted by Ta$_2$O$_5$ and Al in the stack, and will be studied in future works.

Compared with SoA (Table V) of reported SMR, this work presents noticeable frequency scaling along with reasonably good resonator performance metrics, especially the high $Q$ above 200 at 18 GHz. The modest $k^2$ is a combination of using lateral excitation for the device, which has limited available coupling, in addition to having energy leaking into the bragg reflector as seen in Fig 2 (c). It could be further improved using thickness excitation with a bottom electrode, in a slight tradeoff with fabrication complexity.

It is expected that the resonator would have better power handling than a similar IDT laterally excited device on a suspended film. Future studies will focus on reliable large signal analysis to measure power handling capabilities [54].

## V. Conclusion

In this work, we report an SMR at 18.64 GHz using ScAlN on top of $SiO_2/Ta_2O_5$ Bragg reflectors on Si wafer. The resonator shows $k^2$ of 2.0%, high series quality factor $Q_s$ of 156, shunt quality factor $Q_p$ of 142, and maximum Bode quality factor $Q_{max}$ of 210. The device surpasses SoA and highlights the possibility of scaling SMRs toward various front-end signal-processing functions, e.g., filters and oscillators, at FR3 and mmWave bands.

## Acknowledgment

The authors thank the DARPA COFFEE program for funding support and Dr. Ben Griffin, Dr. Todd Bauer, and Dr. Zachary Fishman for helpful discussions.

[1] A. Hagelauer et al., "From Microwave Acoustic Filters to Millimeter-Wave Operation and New Applications," IEEE Journal of Microwaves, vol. 3, no. 1, pp. 484–508, 2023.
[2] R. Lu and S. Gong, "RF acoustic microsystems based on suspended lithium niobate thin films: advances and outlook," Journal of Micromechanics and Microengineering, vol. 31, no. 11, p. 114001, 2021.
[3] S. Gong, R. Lu, Y. Yang, L. Gao, and A. E. Hassanien, "Microwave Acoustic Devices: Recent Advances and Outlook," IEEE Journal of Microwaves, vol. 1, no. 2, pp. 601–609, 2021.
[4] R. Ruby, "A Snapshot in Time: The Future in Filters for Cell Phones," IEEE Microw Mag, vol. 16, no. 7, pp. 46–59, 2015.
[5] R. Aigner, G. Fattinger, M. Schaefer, K. Karnati, R. Rothemund, and F. Dumont, "BAW filters for 5G bands," in 2018 IEEE International Electron Devices Meeting (IEDM), IEEE, 2018, pp. 14–15.
[6] R. V Snyder, G. Macchiarella, S. Bastioli, and C. Tomassoni, "Emerging Trends in Techniques and Technology as Applied to Filter Design," IEEE Journal of Microwaves, vol. 1, no. 1, pp. 317–344, 2021.
[7] R. Aigner, "SAW and BAW technologies for RF filter applications: A review of the relative strengths and weaknesses," in IEEE Ultrasonics Symposium, 2008, pp. 582–589.
[8] R. C. Ruby, P. Bradley, Y. Oshmyansky, A. Chien, and J. D. Larson, "Thin film bulk wave acoustic resonators (FBAR) for wireless applications," in Proceedings of the IEEE Ultrasonics Symposium, 2001.
[9] INC. NTT DOCOMO, "5G Evolution and 6G - White Paper," Distribution, no. January, 2020.
[10] S. Mahon, "The 5G Effect on RF Filter Technologies," IEEE Transactions on Semiconductor Manufacturing, vol. 30, no. 4, 2017.
[11] V. Chulukhadze et al., "Frequency Scaling Millimeter Wave Acoustic Devices in Thin Film Lithium Niobate," in IEEE International Frequency Control Symposium (IFCS), 2023.
[12] J. Kramer et al., "Extracting Acoustic Loss of High-Order Lamb Modes at Millimeter-Wave Using Acoustic Delay Lines," in IEEE MTT-S International Microwave Symposium Digest, 2023.
[13] M. Rinaldi, C. Zuniga, and G. Piazza, "5-10 GHz AlN Contour-Mode Nanoelectromechanical Resonators," in IEEE International Conference on Micro Electro Mechanical Systems, 2009, pp. 916–919.
[14] V. J. Gokhale, M. T. Hardy, D. S. Katzer, and B. P. Downey, "X–Ka Band Epitaxial ScAlN/AlN/NbN/SiC High-Overtone Bulk Acoustic Resonators," IEEE Electron Device Letters, vol. 44, no. 4, pp. 674–677, 2023.
[15] W. Zhao et al., "15-GHz Epitaxial AlN FBARs on SiC Substrates," IEEE Electron Device Letters, p. 1, 2023.
[16] S. Nam, W. Peng, P. Wang, D. Wang, Z. Mi, and A. Mortazawi, "An mm-Wave Trilayer AlN/ScAlN/AlN Higher Order Mode FBAR," IEEE Microwave and Wireless Technology Letters, 2023.
[17] Z. Schaffer, P. Simeoni, and G. Piazza, "33 GHz Overmoded Bulk Acoustic Resonator," IEEE Microwave and Wireless Components Letters, vol. 32, no. 6, pp. 656–659, 2022.
[18] M. Kadota and T. Ogami, "5.4 GHz Lamb wave resonator on LiNbO3 thin crystal plate and its application," Jpn J Appl Phys, vol. 50, no. 7 PART 2, 2011.
[19] Y. Yang, A. Gao, R. Lu, and S. Gong, "5 GHz lithium niobate MEMS resonators with high FoM of 153," in Proceedings of the IEEE International Conference on Micro Electro Mechanical Systems (MEMS), 2017.
[20] J. Kramer et al., "Thin-Film Lithium Niobate Acoustic Resonator with High Q of 237 and k2 of 5.1% at 50.74 GHz," in IEEE International Frequency Control Symposium (IFCS), 2023.
[21] Izhar et al., "A K-Band Bulk Acoustic Wave Resonator Using Periodically Poled Al0.72Sc0.28N," IEEE Electron Device Letters, vol. 44, no. 7, 2023.
[22] S. Nam, W. Peng, P. Wang, D. Wang, Z. Mi, and A. Mortazawi, "A mm-Wave Trilayer AlN/ScAlN/AlN Higher Order Mode FBAR," IEEE Microwave and Wireless Technology Letters, vol. 33, no. 6, 2023.
[23] G. Giribaldi, L. Colombo, P. Simeoni, and M. Rinaldi, "Compact and wideband nanoacoustic pass-band filters for future 5G and 6G cellular radios," Nat Commun, vol. 15, no. 1, 2024.
[24] A. Kochhar et al., "X-band Bulk Acoustic Wave Resonator (XBAW) using Periodically Polarized Piezoelectric Films (P3F)," in 2023 IEEE International Ultrasonics Symposium (IUS), 2023, pp. 1–4.
[25] J. Fang, K. Yang, F. Lin, H. Tao, J. Chen, and C. Zuo, "A Fin-Mounted A5-Mode Lithium Niobate Resonator at 27.58 GHz with k2 of 4.4%, Qp of 448, and FoM of 19.7," in 2024 IEEE/MTT-S International Microwave Symposium - IMS 2024, 2024, pp. 146–149.
[26] J. Kramer et al., "57 GHz Acoustic Resonator with k2 of 7.3 % and Q of 56 in Thin-Film Lithium Niobate," in 2022 International Electron Devices Meeting (IEDM), 2022, pp. 16.4.1-16.4.4.
[27] O. Barrera et al., "Transferred Thin Film Lithium Niobate as Millimeter Wave Acoustic Filter Platforms," in Proceedings of the IEEE International Conference on Micro Electro Mechanical Systems (MEMS), 2024.
[28] S. Cho et al., "23.8-GHz Acoustic Filter in Periodically Poled Piezoelectric Film Lithium Niobate With 1.52-dB IL and 19.4% FBW," IEEE Microwave and Wireless Technology Letters, 2024.
[29] R. Tetro, L. Colombo, W. Gubinelli, G. Giribaldi, and M. Rinaldi, "X-Cut Lithium Niobate S0Mode Resonators for 5G Applications," in Proceedings of the IEEE International Conference on Micro Electro Mechanical Systems (MEMS), 2024.
[30] Cho S et al., "Millimeter Wave Thin-Film Bulk Acoustic Resonator in Sputtered Scandium Aluminum Nitride," Journal of Microelectromechanical Systems, 2023.
[31] G. Giribaldi, L. Colombo, and M. Rinaldi, "6-20 GHz 30% ScAlN Lateral Field-Excited Cross-Sectional Lamé Mode Resonators for Future Mobile RF Front Ends," IEEE Trans Ultrason Ferroelectr Freq Control, vol. 70, no. 10, 2023.
[32] M. Park, J. Wang, D. Wang, Z. Mi, and A. Ansari, "A 19GHz All-Epitaxial Al0.8Sc0.2N Cascaded FBAR for RF Filtering Applications," IEEE Electron Device Letters, p. 1, 2024.
[33] C. C. W. Ruppel, "Acoustic Wave Filter Technology-A Review," IEEE Trans Ultrason Ferroelectr Freq Control, vol. 64, no. 9, 2017.
[34] R. Ruby, "Review and comparison of bulk acoustic wave FBAR, SMR technology," in Proceedings - IEEE Ultrasonics Symposium, 2007.
[35] K. M. Lakin, K. T. McCarron, and R. E. Rose, "Solidly mounted resonators and filters," in Proceedings of the IEEE Ultrasonics Symposium, 1995.
[36] K. Hashimoto, RF bulk acoustic wave filters for communications, vol. 66. 2009.
[37] M. Kadota, F. Yamashita, and S. Tanaka, "9.5 GHz Solidly Mounted Bulk Acoustic Wave Resonator using Third Overtone of Thickness Extension Mode in LiNbO3," in IEEE International Ultrasonics Symposium, IUS, 2022.
[38] L. Lv et al., "BAW Resonator with an Optimized SiO2/Ta2O5 Reflector for 5G Applications," ACS Omega, vol. 7, no. 24, pp. 20994–20999, Jun. 2022.
[39] A. Tag et al., "Next Generation Of BAW: The New Benchmark for RF Acoustic Technologies," in IEEE International Ultrasonics Symposium, IUS, 2022.
[40] Y. Yang et al., "Solidly Mounted Resonators with Ultra-High Operating Frequencies Based on 3R-MoS2 Atomic Flakes," Adv Funct Mater, vol. 33, no. 29, 2023.
[41] T. Kimura, M. Omura, Y. Kishimoto, and K. Hashimoto, "Comparative Study of Acoustic Wave Devices Using Thin Piezoelectric Plates in the 3–5-GHz Range," IEEE Trans Microw Theory Tech, vol. 67, no. 3, pp. 915–921, 2019.

[42] M. Bousquet et al., "Single Crystal LiNbO3 and LiTaO3 Bulk Acoustic Wave Resonator," in IEEE International Ultrasonics Symposium, IUS, 2023.

[43] Z. Schaffer et al., "A Solidly Mounted 55 GHz Overmoded Bulk Acoustic Resonator," in 2023 IEEE International Ultrasonics Symposium (IUS), 2003.

[44] K. Hashimoto, Surface Acoustic Wave Devices in Telecommunications. 2000. doi: 10.1007/978-3-662-04223-6.

[45] T.-H. Hsu, J. Campbell, J. Kramer, S. Cho, M.-H. Li, and R. Lu, "C-Band Lithium Niobate on Silicon Carbide SAW Resonator With Figure-of-Merit of 124 at 6.5 GHz," Journal of Microelectromechanical Systems, pp. 1–6, 2024.

[46] D. F. Urban, O. Ambacher, and C. Elsässer, "First-principles calculation of electroacoustic properties of wurtzite (Al,Sc)N," Phys Rev B, vol. 103, no. 11, 2021.

[47] S. J. Orfanidis, Electromagnetic Waves and Antennas, vol. 2, no. Rutgers U. 2004. doi: 10.1016/B978-075064947-6/50011-3.

[48] L. Colombo, G. Giribaldi, and M. Rinaldi, "18 GHz Microacoustic ScAlN Lamb Wave Resonators for Ku Band Applications," in 2024 IEEE International Microwave Filter Workshop (IMFW), 2024, pp. 84–86.

[49] M. Scherer, J. Pistner, and W. Lehnert, "UV- and VIS filter coatings by plasma assisted reactive magnetron sputtering (PARMS)," in Optics InfoBase Conference Papers, 2010.

[50] A. Zoeller, M. Boos, R. Goetzelmann, H. Hagedorn, and W. Klug, "Substantial progress in optical monitoring by intermittent measurement technique," in Advances in Optical Thin Films II, 2005.

[51] M. Akiyama, T. Kamohara, K. Kano, A. Teshigahara, Y. Takeuchi, and N. Kawahara, "Enhancement of piezoelectric response in scandium aluminum nitride alloy thin films prepared by dual reactive cosputtering," Advanced Materials, vol. 21, no. 5, 2009.

[52] D. A. Feld, R. Parker, R. Ruby, P. Bradley, and S. Dong, "After 60 years: A new formula for computing quality factor is warranted," in Proceedings - IEEE Ultrasonics Symposium, 2008.

[53] C. M. Lin et al., "Temperature-compensated aluminum nitride lamb wave resonators," in IEEE Transactions on Ultrasonics, Ferroelectrics, and Frequency Control, 2010.

[54] J. Verspe, "Large-signal network analysis," IEEE Microwave Magazine, vol. 6, no. 4. 2005. doi: 10.1109/MMW.2005.1580340.